# Calculation of the temporal structure for a pulsed slow positron beam based on superconducting accelerator

H. Q. Zhang [a,b,1], P. Kuang [b,1], Y. H. Hu [b], F. Y. Liu [b], X. Z. Cao [a,b,*], B. Y. Wang [b], X. P. Li [b,*]

[a] *Center for High Energy Physics, Henan Academy of Sciences, Zhengzhou 450046, China*
[b] *Institute of High Energy Physics, Chinese Academy of Sciences, Beijing 100049, China*
*E-mail: caoxzh@ihep.ac.cn, lxp@ihep.ac.cn*

**Abstract:**

The development of superconducting accelerator (SCA) provides a novel approach to generating pulsed slow positron beams with high time resolution and high intensity. SCAs can produce electron beams with repetition frequencies on the order of MHz and pulse widths of less than 100 ps. A pulsed positron beam based on SCA can, to a certain extent, preserve the excellent temporal structure of the primary electron beam. However, thermalization, diffusion, and surface re-emission of positrons in the moderator will inevitably lead to time broadening of positron pulses. In this paper, a calculation model coupling Geant4 Monte Carlo simulations with positron diffusion theory is established to evaluate the effect of the positron moderation process on time broadening. After being moderated by a tungsten foil, the initial time broadening of the pulsed slow positron beam is approximately 320 ps. By combining the time broadening calculation of pulsed beam during its transport, it is expected that the pulsed slow positron beam generated by the SCA, can be directly applied to the measurement of positron annihilation lifetime.

## 1 Introduction

The measurement of positron annihilation lifetime (PAL) based on pulsed slow positron beams is a powerful tool for characterizing open volume defects, pores and free volume in the surface or thin film materials [1, 2]. When positrons are injected into the material, their lifetime depends on the electron density at the specific location. In defects, the positron lifetime can be extended to the order of hundreds of picoseconds or even nanoseconds. By measuring the PAL spectrum, microscopic structural information such as type and concentration of defects, and size distribution of pores or free volume can be obtained.

Most slow positron beam devices use $^{22}$Na radioactive isotope as the positron source, and the counting rate is limited by the source activity. The intensity of the positron source can be increased by $10^{2\text{-}3}$ times through ($\gamma$, $e^+e^-$) reaction. Numerous slow positron beam devices have been well constructed based on nuclear reactors (such as the NEPOMUC source [3], PULSTAR source [4], etc.) or electron linear accelerators (such as the positron source at KEK [5], AIST [6], etc.). However, these positron beams cannot be directly used for PAL measurements. Since the positron beams based on $^{22}$Na and nuclear reactor are continuous without any temporal structure information, they cannot provide the starting time signal required for PAL measurements. The initial temporal structure of positron sources based on conventional LINACs (pulse width ~μs, frequency <1 kHz) also does not meet the timing requirements. As shown in Fig. 1(a), for this type of positron beam, a Penning trap is usually used to sufficiently stretch the pulse into a continuous beam. Then, it is generally cut into pulsed beams with a repetition frequency of several tens of MHz and a width of 2-5 ns by a chopper. Finally, the positron pulse will be further compressed by bunchers to obtain a positron pulse of less than 350 ps at the sample target [7-9]. The entire system is highly complex, and the manipulation on the temporal structure of beam results in a loss of nearly 90% of the beam intensity, depending on the duty cycle of the chopper, bunching efficiency and transport efficiency.

---

* Corresponding author.

[1]These authors contributed equally to this work.

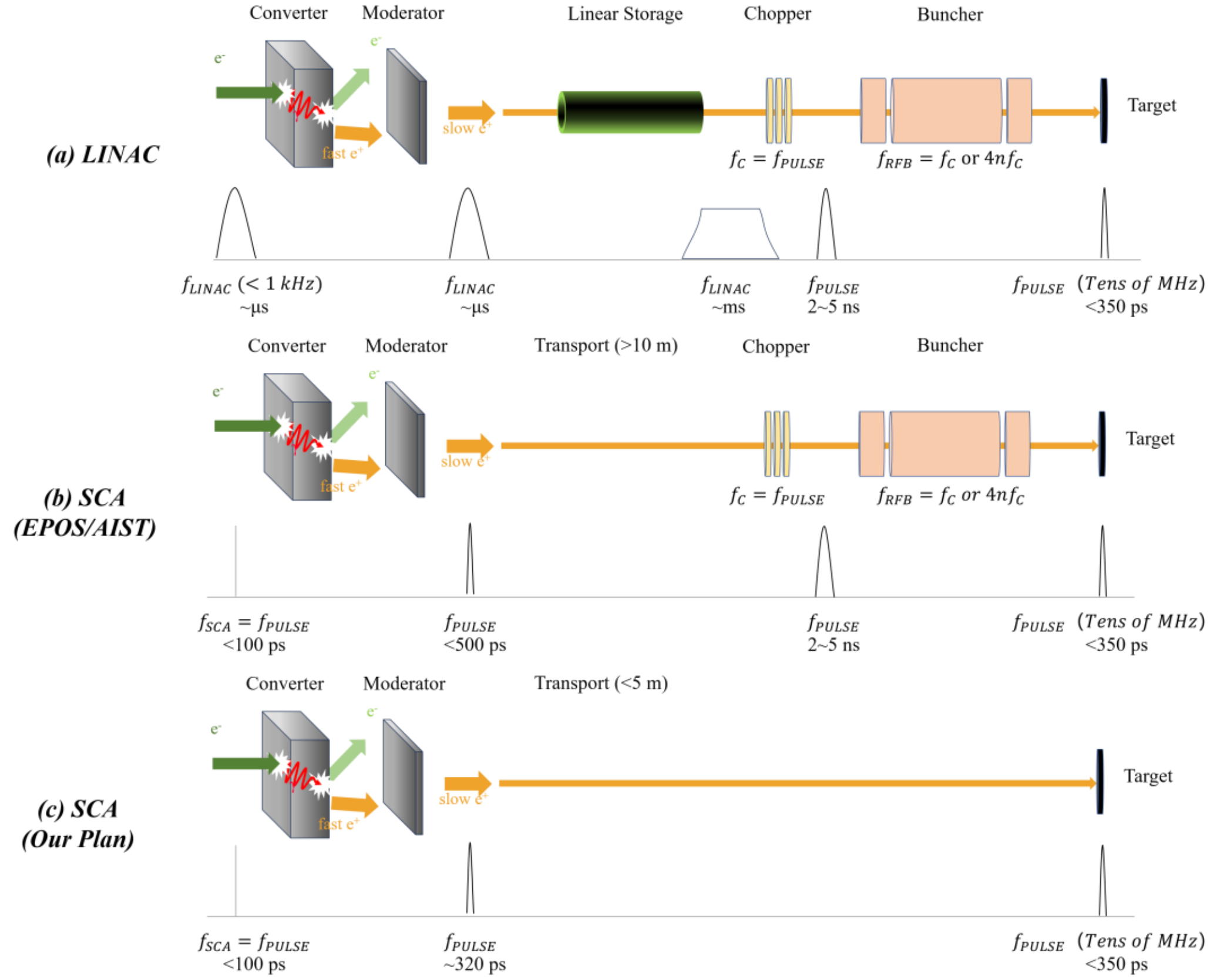

**FIG. 1.** Schematic diagram of the scheme for generating pulsed slow positron beams based on electron accelerators. (a) is based on a conventional LINAC; (b) is based on SCA for EPOS and AIST; (c) is our proposed plan based on SCA for PAPS.

The SCA technology has been developed to generate electron beams with high repetition frequency (~MHz) and short pulse width (<100 ps), which provides a new solution to address the above challenges. For instance, the ELBE radiation source at the Research Centre Dresden-Rossendorf is equipped with a similar SCA (40 MeV, 13 MHz, 5 ps). Due to its unique timing properties, the first bunched positron beam based on SCA, EPOS, has been built [10]. At AIST, a scheme for pulsed positron beam generation has been proposed based on two SCA modules (15 MeV, 500 MHz, <100 ps), planning to replace the original scheme based on a conventional 70 MeV LINAC [11]. The scheme is shown in Fig. 1(b), which does not require the linear storage device. However, the chopper and buncher are still needed to reduce the time broadening due to the long-distance transport process. Ideally, the pulsed positron beams can be directly used for PAL measurements without additional time structure manipulation, thereby greatly increasing utilization efficiency of positrons. This goal is planned to be achieved through the SCA of the Platform of Advanced Photon

Source Technology R&D (PAPS), as shown in Fig. 1(c). The beam test system located at PAPS is used to test the key technologies of the 650 MHz superconducting radio frequency system, which is based on a photocathode direct current electron gun and two 650 MHz 2-cell superconducting cavities [12]. The basic parameters are shown in Table I. This SCA module can provide a maximum acceleration gradient of 10 MV/m, and the energy of the electron beam can be adjusted between 5-10 MeV. The electron beam with lower energy can be used to hit target to generate positrons. Although the high-energy electron beam can obtain a high probability of positron production, it will impose higher requirements on biological shielding [13]. Low-energy (≤ 10 MeV) accelerators produce very few neutrons, significantly reducing the difficulty of radiation shielding, which is an advantage for maintaining sufficient compactness to reduce the time broadening due to beam transport [14, 15].

The SCA has brought revolutionary improvements to the temporal structure of positron sources. However, how large the time broadening caused by moderation is has so far lacked systematic evaluation. In this paper, by establishing a coupled model of Geant4 Monte Carlo simulation and positron diffusion theory, the time distribution of slow positrons emitted from the surface of moderator is calculated. After the beam transport process, the feasibility of directly applying the pulsed slow positron beam generated by a SCA to the PAL measurement is also evaluated.

**TABLE I.** Beam parameters of SCA for PAPS.

| Parameters | Value |
|---|---|
| High voltage of electron gun | 350-500 kV |
| Maximum acceleration gradient | 10 MV/m |
| Electron beam energy | 5-10 MeV (Adjustable) |
| Average beam current | 0.1-3 mA (Adjustable) |
| Emittance | <2 mm·mrad |
| Bunch charge | 15.4 pC (Adjustable) |
| Bunch repetition frequency | 650 MHz (Adjustable) |
| Bunch length | 2-20 ps (Adjustable) |

**2 Physical model**

The generation of pulsed slow positron beams based on superconducting accelerators can be divided into three stages: the production of high-energy positrons, the moderation of positrons, and the transport of the slow positron beam.

Positrons are generated by pair production with the Coulomb field of the target nucleus when high-energy electron beams hit a high-Z metallic target (such as W or Ta). This process is almost instantaneous and does not introduce any additional time broadening. The time scale of this process is determined by the initial electron pulse.

These positrons have a relatively wide and continuous energy distribution ranging from zero to several MeV. They can be used for slow positron beam after being slowed down by foils [16] or meshes [14] made of materials with negative work functions (such as W or Pt). After entering a moderator, the high-energy positrons gradually lose energy through inelastic scattering with phonons and electrons, and eventually thermalize to near the thermal energy. This process is called thermalization, and its time scale is on the order of picoseconds. Dryzek obtained from Geant4 simulations that positrons emitted by a $^{22}$Na source (with an average energy of 216 keV) and implanted into tungsten have an average thermalization time of approximately 0.44 ps, with the time distribution extending up to 1.2 ps [17]. Brandt has publicized an approximate formula for the relationship between thermalization time, $t_{th}$, and incident positron energy, $E_i$, which is $t_{th} \propto E_i^{1.2}$ [18]. Therefore, it can be expected that thermalization time of the positrons with higher average energy (~2 MeV) produced by the electron accelerator, will be further extended. The positrons that complete the thermalization process either diffuse to the surface of the moderator or annihilate with the orbiting electrons within the moderator. Generally, positrons that can diffuse to the surface and be emitted from the surface are called slow positrons. The difference in implantation depth will directly lead to the dispersion of the time taken for diffusion to the surface. The diffusion time of positrons in a perfect lattice is determined by the annihilation lifetime. For tungsten, the lifetime in lattice is approximately between 100 ps and 110 ps [19]. Fig. 2 shows the time distribution of positrons that thermalized at $t_0$ and diffuse to the surface of the transmission-type W foil moderator. The finite lifetime of positrons means that even if all the positrons are implanted simultaneously and with the same implantation depth, the probability of their surface emission decreases exponentially over time, with the time constant being equal to the lifetime. Therefore, the width of the positron pulse will at least extend to the order of the lifetime. Furthermore, the diffusion of positrons within the lattice is influenced by lattice defects, impurities, and temperature. The diffusion paths and effective diffusion coefficients of different positrons may vary, which could further increase the time dispersion [20]. In summary, the contribution of the positron moderation process to the time broadening ($\Delta t_S$) mainly comes from three aspects: the difference in implantation depth, the finite lifetime of positron, and the statistical fluctuation of the diffusion coefficient.

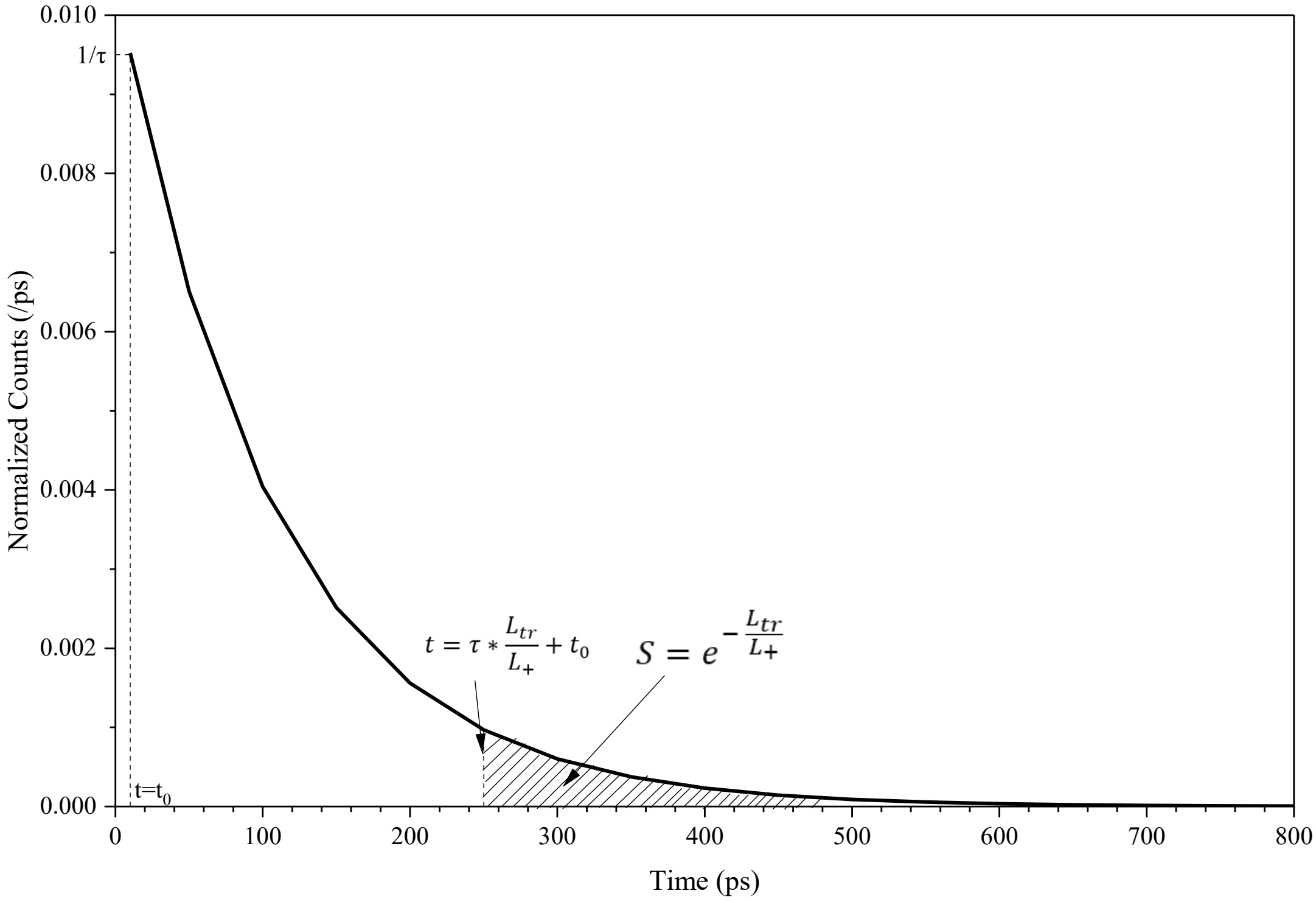

**FIG. 2.** The time distribution of positrons that thermalized at $t_0$ and diffuse to the surface of the transmission-type W foil moderator. The black solid line represents the time distribution of positrons that thermalized at the transmission surface of the moderator ($z = D$) and at $t_0$. The area of the shaded region ($S = \exp(-L_{tr}/L_+)$) represents the probability integral of the time distribution of positrons that thermalized at the any distance, $L_{tr}$, from transmission surface ($L_{tr} = D - z$) and at

$t_0$. The lifetime for W, $\tau$, is set to 105 ps.

The key issue is how to calculate the contribution of the moderation process to the time broadening of a pulsed positron beam. Thermalized positrons diffuse through the moderator. In a perfect lattice, statistical fluctuations of the diffusion coefficient can be neglected, and the diffusion process is determined by the positron diffusion length, $L_+$,

$$L_+ = \sqrt{\tau D_+} \tag{1}$$

where $\tau$ is lattice lifetime and $D_+$ is the thermal diffusion coefficient. In this paper, $D_+$ for W is set as 1.26 $cm^2/s$ [21]. The Geant4 Monte Carlo simulation is a mature tool for simulating the transport of high-energy particles, which has been widely used in positron source simulations [22]. Geant4 is typically applicable for the transport simulation of keV-MeV particles, but is no longer suitable for processes such as coherent diffusion of ~eV positrons in the lattice potential field. Therefore, Geant4 cannot simulate the complete diffusion dynamics process of positrons after thermalization. The macroscopic diffusion theory must be introduced to describe the diffusion behavior of positrons in the moderator. To better understand the diffusion process of positrons, in this paper, an isotropic transmission-type tungsten thin foil is adopted as the moderator. The diffusion behavior of positrons in the moderator is described by the following time-dependent diffusion equation [23]:

$$\frac{\partial C(z,t)}{\partial t} = D_+ \frac{\partial^2 C(z,t)}{\partial z^2} - \frac{C(z,t)}{\tau} + f(z,t_0) \tag{2}$$

The thickness of the moderator foil is usually μm to tens of μm, which is much smaller than the horizontal dimension. Therefore, in this paper, a one-dimensional diffusion approximation is adopted, and $C(z,t)$ is defined as the one-dimensional linear density, which is the probability of positrons at a unit depth. $f(z,t_0)$ represents the implantation probability profile of fast positrons emitted from the converter, indicating the probability of implantation within the depth range of $(z, z+dz)$ and during the time interval of $(t_0, t_0 + dt_0)$. To establish an equation that can be solved analytically, the following assumptions are made:

i. The moderator is an isotropic and homogeneous thin foil with a thickness of D;

ii. A fully absorbing boundary is adopted and the positrons that diffuse to the surface are immediately emitted;

iii. At the initial moment ($t = t_0$), the implantation distribution $f(z,t_0)$ acts instantaneously on the moderator, and the initial implantation time $t_0$ only introduces an overall time delay.

For the Green's function solution of non-time-dependent diffusion equations, it is approximately given by the positron current, $J(z)$, passing through the surface at $z = D$ [24]:

$$J(z) = \exp(-\frac{L_{tr}}{L_+}) \tag{3}$$

where $L_{tr} = D - z$ represents the distance from the transmission surface ($z = D$) of the moderator. $J(z)$ is called the moderator efficiency kernel and indicates the probability that a thermalized positron diffuses to the transmission surface. $J(z)$ can be convolved with the implantation profile $f(z,t_0)$ to obtain the fraction of thermalized positrons diffusing to the transmission surface. The moderation efficiency is defined as the ratio of the number of slow positrons emitted from the surface of the moderator to the number of fast positrons emitted from the converter. Therefore, the moderation efficiency, $\varepsilon$, can be expressed as:

$$\varepsilon = y_0 \int_{-\infty}^{+\infty} \int_0^D f(z,t_0) \exp(-\frac{L_{tr}}{L_+})\, dz dt_0 \tag{4}$$

where $y_0 = 0.33$ is the branching ratio of thermal positron emission from the surface of the moderator to the vacuum [25]. For time-dependent diffusion equations, except for a few special cases, there are no direct analytical solutions. Britton has already discussed in detail the analytical solutions for time-dependent diffusion equations with initial Gaussian derivative distribution and Dirichlet boundary conditions [26]. For the time-dependent diffusion equations with arbitrary implantation profiles, the more general solution approach is usually the numerical solution method, such as the simplest finite difference method [23]. In this paper, the approximate solutions of time-dependent diffusion equations will be constructed directly based on the solutions of non-time-dependent diffusion equations and the probability distribution of surface emission. The black solid line in Fig. 2 represents the time distribution of positrons that thermalized at the transmission surface of the moderator $z = D$ and at $t_0$. It follows an exponential decay distribution, which is expressed as $\exp(-(t - t_0)/\tau)/\tau$, and its integral value is 1. Similarly, the time distribution of the thermalized positrons diffusing from any position $z$ to the transmission surface should also satisfy an exponential decay distribution. The integral value of their time distribution should be the value of the moderator efficiency kernel, $\exp(-\frac{L_{tr}}{L_+})$, as shown in the shaded region of Fig. 2. Therefore, at any time, $t_0$, and at any position $z$, the probability distribution of time of the thermalized positrons diffusing to the transmission surface can be expressed by $T_f(z,t)$:

$$T_f(z,t) = \begin{cases} 0 & , t < \tau * \frac{L_{tr}}{L_+} + t_0 \\ \frac{1}{\tau} * \exp(-\frac{t-t_0}{\tau}), & t \geq \tau * \frac{L_{tr}}{L_+} + t_0 \end{cases} \tag{5}$$

where $T_f(z,t)$ is called time distribution kernel. $T_f(z,t)$ can be convolved with $f(z,t_0)$to obtain the total time probability distribution, T(t):

$$T(t) = \frac{y_0 \int_{-\infty}^{+\infty} \int_0^D f(z,t_0) T_f(z,t) dz dt_0}{\varepsilon} \tag{6}$$

$T(t)$ has been normalized. The broadening of $T(t)$ is described by its full width at half maximum (FWHM), and its value is expressed as $\Delta t_S$.

The positrons that diffuse to the surface of the moderator can overcome the surface potential barrier and be emitted from the surface into the vacuum with an energy of several eV, depending on the work function of the moderator material [27]. The slow positrons can be electrostatically accelerated and transported to the sample target by a confining magnetic field. During the transport process, due to the finite energy spread of slow positron beam, the positron pulse will have a time spread. Assuming that the energy, $E_0$, and energy spread, $\Delta E$, of any positron at any phase of the positron pulse are the same, and $\Delta E \ll E_0$, the time broadening, $\Delta t_T$, caused by a certain phase of the positron during the transport process at a distance of $L$ can be expressed by the following formula [28]:

$$\Delta t_T = \sqrt{\frac{m_e}{8}} \cdot \frac{L}{E_0^{3/2}} \cdot \Delta E \tag{7}$$

where $m_e$ is the rest mass of an electron. Therefore, the time broadening of the pulsed positron beam, Δt, is obtained by the convolution of the contribution from the moderator process ($\Delta t_S$) and the contribution from the transport process ($\Delta t_T$), that is:

$$\Delta t = \sqrt{\Delta t_S{}^2 + \Delta t_T{}^2} \tag{8}$$

In this paper, a simplified model of electron sources, converter, and moderator is developed in Geant4. A one-way electron beam source is defined, which is emitted from the surface of the converter and vertically towards the target. The number of emitted particles is $10^9$. The electron beam energy is 10 MeV, which is the maximum energy provided by the SCA module of PAPS. To study the effect of pulse width of the electron beam on the time broadening of slow positrons, the time distribution of electron beam is set as Gaussian distribution, with 6σ values of 0 ps, 2 ps, 20 ps, 100 ps, 200 ps, 300 ps, 500 ps, 700 ps, and 1000 ps, although in reality the pulse width may not exceed 100 ps. For the SCA of PAPS, the pulse width of the electron beam does not exceed 20 ps. In this paper, the time broadening of the electron beam, $\Delta t_E$, is defined as the FWHM (2.355σ) of the Gaussian distribution. The beam spot is a Gaussian distribution with a radial standard deviation of 2 mm, $\sigma_r$. To avoid edge effects, the radii of the converter and moderator are set to 20 mm. The material of the converter is tungsten and its thickness is 1.5 mm. This thickness has been optimized to obtain the maximum probability of fast positrons emitted from the converter, which is approximately $2\times10^{-3}$. A moderator foil with a thickness of 10 μm is placed 1 mm behind the converter. When calculating the depth profile of positron implantation, once the energy of the positrons falls below 0.1 keV, they are regarded as "stopping". The stopping profile is almost identical to the implantation profile produced by the physical thermal energy cutoff [24]. Therefore, the calculation process for the moderation efficiency and the time distribution of slow positrons is as follows: (1) obtaining the positron implantation depth profile $f(z, t_0)$ by Geant4 simulation, which serves as the source term for the diffusion process; (2) based on diffusion theory, calculating the moderation efficiency kernel, $J(z)$, and the time distribution kernel $T_f(z, t)$ for thermalized positrons at any position, $z$, and any time, $t_0$, diffusing to the transmission surface; (3) respectively performing numerical convolution of $J(z)$ and $T_f(z, t)$ with the $f(z, t_0)$ to obtain the positron moderation efficiency, $\varepsilon$, and the total time probability distribution, $T(t)$. In this paper, an explicit functional form of $f(z, t_0)$ is not required; instead, a statistical sampling method is employed, where a large number of discrete thermalization events $f(z, t_0)$ are generated by Monte Carlo simulation and used as the source term, each subsequently processed individually using diffusion theory.

## 3 Results and Discussion

Fig. 3 shows the implantation depth profile, $f(z)$, calculated by Geant4 when the electron beam is emitted simultaneously at $t = 0$. The time integral of $f(z, t_0)$ has been performed. The thermalization depth of high-energy positrons in the moderator is not uniform. The implantation profile of monoenergetic positrons follows a nearly exponential decay distribution, which can be theoretically expressed by the Makhovian formula depending on the incident energy and incident angle [24]. For the positron beam obtained by a 10 MeV electron beam hitting the target, it has a continuous energy spectrum ranging from 0 to 9 MeV with a certain degree of distribution, so the implantation depth has a more complex distribution. Since the incidence process of the electron beam and the generation process of the positron are independent, the implantation depth profile obtained by performing the time integral of $f(z, t_0)$ for electron beams with different widths should be the same, as shown in Fig. 3. By convolving $f(z)$ with the moderator efficiency kernel, $\exp(-\frac{L_{tr}}{L_+})$ and then scaling by the re-emission branching ratio, $y_0$, the moderator efficiency ε can be calculated, which is approximately $1\times10^{-5}$ and is independent of the electron beam time broadening.

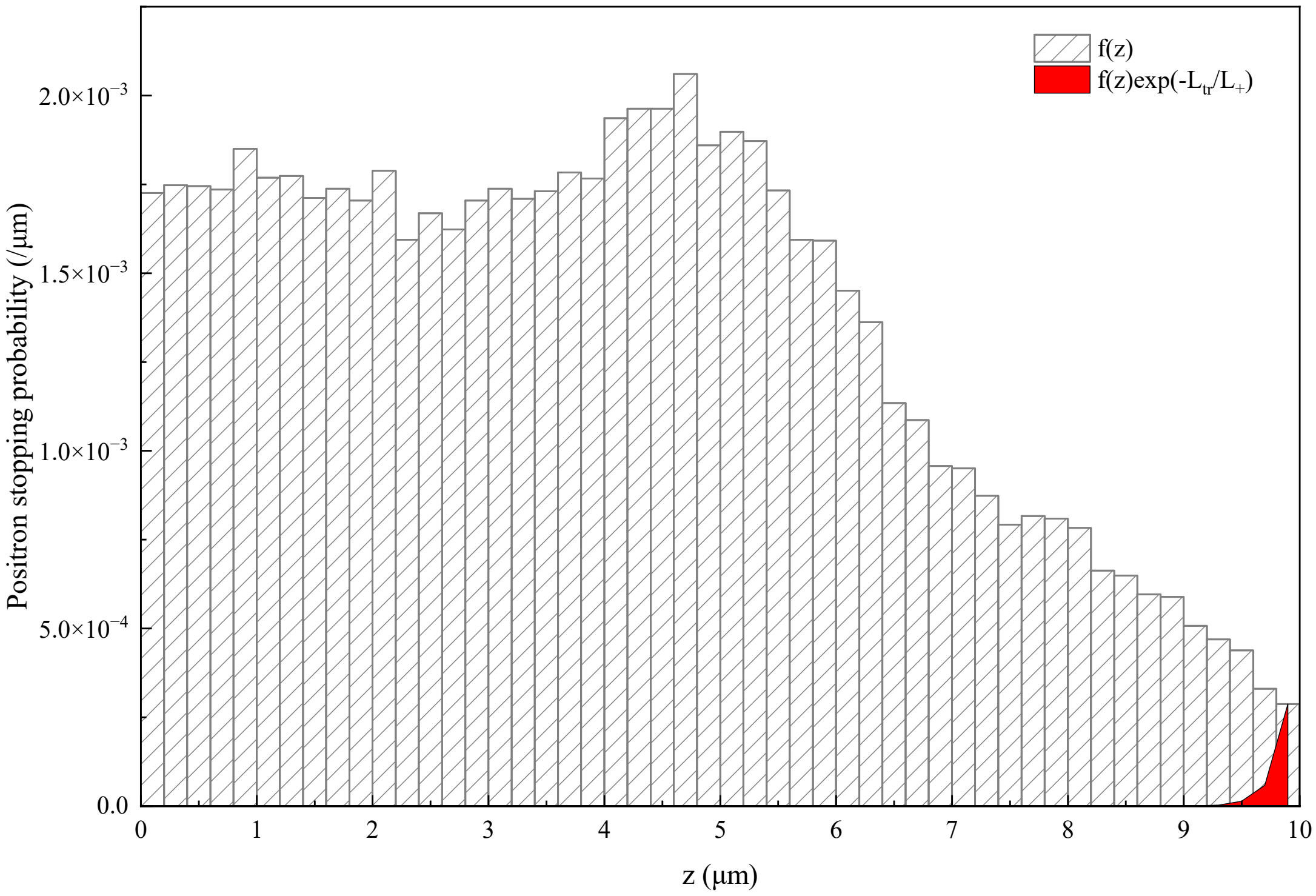

**FIG. 3.** The implantation depth profile, $f(z)$ of positrons in moderator with a thickness of 10 μm calculated by Geant4. The total area of the red region represents the total probability that the positrons diffuse to the transmission surface of the moderator. The electron beam is emitted simultaneously at $t = 0$. The width of the divided intervals is 0.2 μm.

Fig. 4 shows that the time broadening of fast positrons, thermalized positrons, and slow positrons varies with the time broadening of the initial electron beam. When the electron beam hits the target without time broadening, the results indicate that the time broadening of fast positrons, thermalized positrons, and slow positrons are 7 ps, 11 ps, and 317 ps, respectively. Thus, the pair production and the thermalization process hardly add additional time broadening. The time broadening of fast positrons and thermalized positrons is entirely dependent on the time performance of the initial electron beam. When time broadening of the electron beam is less than 80 ps, the broadening caused by the moderation process plays a decisive role. Therefore, for the electron beam pulse widths of interest (2-20 ps for PAPS), the contribution of the initial electron beam to the time broadening of slow positrons can be neglected, and only the contribution of the moderator process ($\Delta t_S$~320 ps) should be considered. The solid lines in Fig. 4 indicate that the time distributions of fast positrons, thermalized positrons, and slow positrons can be regarded as the convolution of the time distribution of positrons obtained when the electron beam has no broadening and the time distribution of initial electron beam. Therefore, it can be assumed that the time distributions are mutually independent Gaussian distributions, and the time broadening of positrons can be directly estimated through convolution.

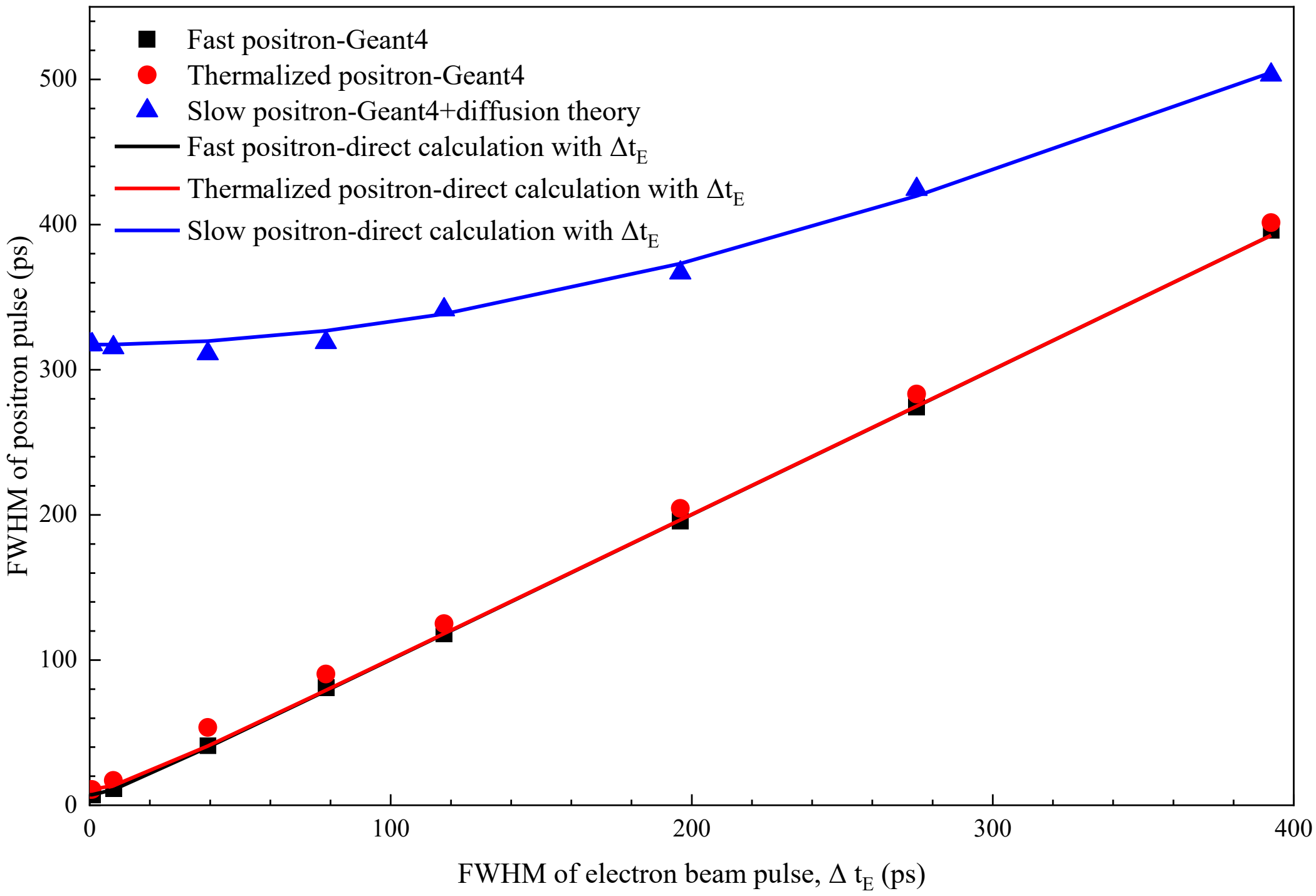

**FIG. 4.** The time broadening of fast positrons, thermalized positrons, and slow positrons varies with the time broadening of the initial electron beam. The solid points represent the time broadening obtained by Geant4 simulation. The time broadening of slow positrons is calculated by combining the diffusion theory. The solid line represents the broadening of time distribution obtained by convolving the time distribution of the positrons calculated when the electron beam has no broadening, with the time distribution of the initial electron beam. The values of time broadening for fast positrons, thermalized positrons, and slow positrons are calculated by $\sqrt{(7\,ps)^2+{\Delta t_E}^2}$、$\sqrt{(11\,ps)^2+{\Delta t_E}^2}$、$\sqrt{(317\,ps)^2+{\Delta t_E}^2}$, respectively. It is assumed that the distributions are all Gaussian distributions.

In fact, slow positrons beam pulse is not strictly Gaussian distribution. Fig. 5 shows the time distribution of pulses of slow positron beams, calculated by the "Geant4 + diffusion theory" method. Each distribution has been normalized by dividing by the moderation efficiency, $\varepsilon$. Due to the time broadening of the electron beam, the calculated time distribution of pulsed slow positrons will no longer be smooth, which becomes more obvious at larger time broadening (6σ > 200 ps, i.e., FWHM > 78.5 ps). This is caused by the sample size effect. In Monte Carlo simulations, it is usually possible to reduce the data variance by increasing the number of simulated particles. However, this was not done in this paper, because the curves obtained within the electron-beam time-broadening range of interest (6σ < 20 ps) were already smooth enough. The inset in Fig. 5 shows the time distribution of slow positron pulse when the electron beam has no time broadening. The rising edge of the slow positron pulse (from 10% to 90% rise time) is approximately 100 ps, and is mainly determined by the implantation time, $t_0$, and the diffusion broadening. The pulse tail approximately follows an exponential decay, determined by the lifetime of the positron in the moderator. The duration of the tail (from 10% to 90% fall time) is approximately 400 ps.

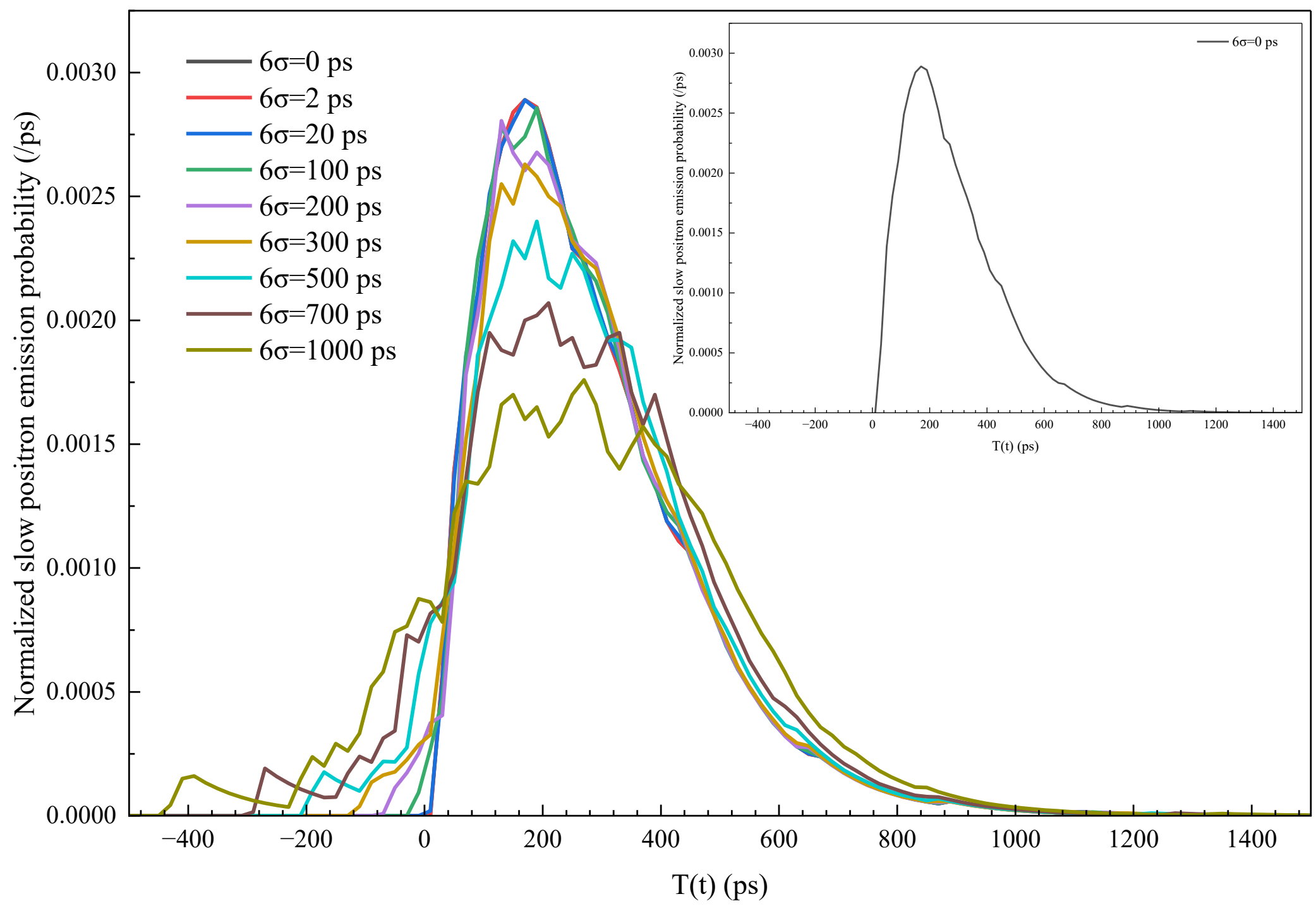

**FIG. 5.** The time distribution of pulses of slow positron beams, calculated by the "Geant4 + diffusion theory" method when the incident electron beam has different time broadening. The inset shows the time distribution of slow positron pulse when the electron beam has no time broadening. The width of the divided intervals is 20 ps.

The final time broadening of the positron pulse, $\Delta t$, must also include the contribution of the transmission process, $\Delta t_T$. $\Delta t_T$ is proportional to the initial energy spread of the slow positrons. The energy spread of slow positrons depends on the geometry and surface quality of the moderator. Theoretically, the energy spread can be reduced to near the thermal energy by using a tungsten foil moderator with a clean surface. In practical use, the energy spread of positrons for tungsten foil moderator is generally 0.5 eV [27, 29]. Due to the differences in annealing treatment time and treatment method of the moderator, the energy spread of positrons can potentially broaden to the order of the positron work function (~3 eV). $\Delta t_T$ is proportional to the transport distance, $L$ and inversely proportional to $E^{3/2}$. In the EPOS scheme, the 2 keV positrons are transmitted to the material research target located approximately 12 m away from the moderator and the time broadening of the positron pulse is 1.5 ns without beam bunching [10]. In the beam line of AIST, the total distance from the moderator to the sub-harmonic buncher is around 18 m, and the pulse width can be ensured to remain below 5 ns with a transport energy of several hundred eV, which is required for the sub-harmonic buncher [11]. In our scheme, the use of a 10 MeV electron beam can significantly reduce the difficulty of radiation shielding and maintain sufficient compactness. In the preliminary design, it is expected that the transport distance will be less than 5 m. Fig. 6 shows the time broadening of positron pulse, $\Delta t_T$, as a function of positron energy, $E_0$, during the transport process. If the time broadening of positron pulse at the sample target is expected to be less than 350 ps, $\Delta t_T$ needs to be less than 140 ps. Even if the initial energy spread of the positrons is 3 eV, the goal can be achieved by simply increasing the positron transport energy to more than 2 keV. Although high-energy transport faces some engineering challenges, it can be fully realized. Firstly, if the positron beam travels through a curved magnetic field, due to the drift effect, there will be a lateral displacement perpendicular to the curved plane. However, this displacement can be eliminated by applying a correction magnetic field perpendicular to the bending plane. Secondly, increasing the transport energy means applying a higher accelerating electric field at the exit of the moderator, which may introduce additional energy spread. However, for a typical DC acceleration structure, the increase in energy spread could be controlled within an acceptable range when accelerating positrons from a few eV to 2 keV.

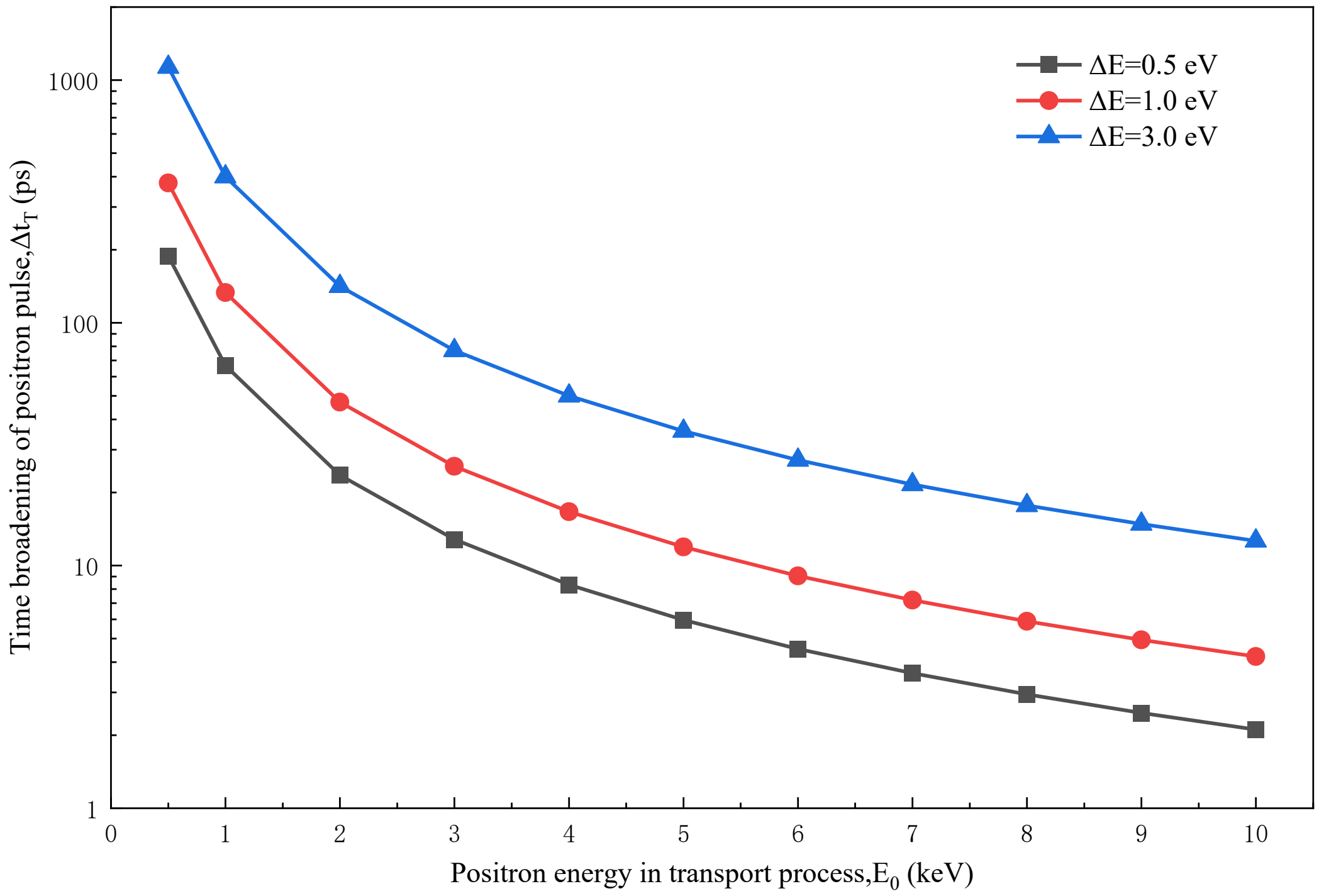

**FIG. 6.** The time broadening of positron pulse, $\Delta t_T$, as a function of positron energy, $E_0$, during the transport process. The transport distance is fixed at 5 m. The initial energy spread of positron, $\Delta E$, is set to 0.5 eV, 1.0 eV, and 3.0 eV respectively.

When using pulsed slow positron beams for PAL spectrum measurements, it is necessary to ensure that the pulse repetition period is long enough to guarantee that all positrons from the previous pulse have annihilated, avoiding spectra disturbance. Existing studies have shown that in order to clearly separate the lifetime components from the spectrum, the pulse repetition period must be at least 5 times the longest lifetime value [30]. If the repetition frequency of the SCA for PAPS is not adjusted in any way, the maximum measurable lifetime is about 300 ps, which is only suitable for the study of open volume defects in most metals [31] and semiconductors [1] with at most a few vacancies. When characterizing pores with a size larger than 1 nm for porous materials and polymer materials, the positron lifetime extends from a few ns to nearly 142 ns (the lifetime of o-Ps in vacuum) [32]. In principle, the repetition frequency of the SCA can be adjusted to other values to obtain an appropriate time window for PAL measurement. It is planned to cut one pulse out of every 32 pulses from the SCA pulse sequence, at which point the repetition period is approximately 50 ns, allowing the measurement of the longest lifetime of about 10 ns. This covers most applications of interest. Compared to the high-energy scheme, the probability of positron production for 10 MeV electron beam is relatively lower. However, the SCA system of PAPS can provide a maximum average electron beam current of 3 mA.

It is preliminarily estimated that the beam intensity is about $3.8 \times 10^8$ $e^+$/s at full power operation. In the pulse selection mode, the slow positron beam intensity can also reach the order of $10^7$ $e^+$/s. This scheme is expected to be directly applied to PAL measurement with both high time resolution and high counting rate.

## 4 Conclusion

In this paper, a coupled computational model combining Geant4 Monte Carlo simulation with positron diffusion theory is established to evaluate temporal structure for pulsed slow positron beams based on SCAs. The effects of the moderation and beam transport process on time broadening of the positron pulses are systematically studied. By using a tungsten foil moderator with a thickness of 10 μm, the initial time broadening of slow positron emitted from the transmission surface is approximately 320 ps. Through short distances (<5 m) and high energy (~2 keV) transport, it is expected to obtain a pulsed slow positron beam with time resolution better than 350 ps and beam intensity of $10^7$ $e^+$/s without additional beam bunching systems. By flexibly adjusting the pulse repetition frequency, this plan has the potential to be directly applied to PAL measurement, especially for the open-volume defect study in metals and semiconductors, and pores in short-lifetime (<10 ns) polymer systems. To achieve this goal in engineering, it is necessary to continue detailed research. The following are the existing limitations and the subsequent work that needs to be carried out:

i. This model assumes that the moderator is an ideal homogeneous medium and does not consider the effects of actual factors such as lattice defects and surface oxidation on positron diffusion and surface emission. Although these issues can be mitigated by sufficient annealing of the moderator in engineering, a diffusion model including defect states can be introduced for calculation to further evaluate the impact.

ii. The actual transport process of the beam in the electro-magnetic field is not considered at present. Dynamics simulations of beam transport (such as SIMION simulations) will be carried out to further verify the feasibility of high-energy transport.

iii. The engineering design the entire system is currently underway.

## Acknowledgments

This work was supported in part by the Scientific Research and Development Project of Henan Academy of Sciences (No.241838241), Innovation Program of the Institute of High Energy of Chinese Academy of Sciences (No.2024000078), and Youth Innovation Promotion Association of Chinese Academy of Sciences (No.2023016).